\documentclass{pasj01}
\usepackage{mathrsfs} 
\usepackage{graphicx} 
\usepackage{mediabb}
\usepackage{color}  
  
\title{Focusing MHD Wave as a Trigger of Star Formation}  
\author{Yoshiaki \textsc{Sofue}\altaffilmark{} }
\altaffiltext{}{Institute of Astronomy, The University of Tokyo, Mitaka, Tokyo 181-0015}


\begin{document} 
\date{ } 
\maketitle      
 
\def\vlsr{v_{\rm LSR}} \def\Msun{M_\odot} \def\deg{^\circ} \def\/{\over}\def\kms{km s$^{-1}$} \def\Tb{T_{\rm B}} \def\sin{{\rm sin}\ } \def\cos{{\rm cos}\ } \def\Hcc{ H cm$^{-3}$ } \def\co{$^{12}$CO$(J=1-0)$ } \def\Htwo{H$_2$ } \def\be{\begin{equation}} \def\ee{\end{equation}} \def\({\left(} \def\){\right)} \def\[{\left[} \def\]{\right]}\def\Ico{I_{\rm CO} } \def\sech{{\rm sech}} \def\Htwocc{${\rm H_2\ cm^{-3}}$} \def\pr{p_r}\def\pt{p_\theta}\def\pp{p_\phi}\def\d{\partial}\def\cot{{\rm cot}} \def\Alf{Alfv{\'e}n } \def\muG{$\mu$G } \def\nH{n_{\rm H}}
\def\Vu{V_{\rm unit}} \def\rhou{\rho_{\rm unit}} \def\Bu{B_{\rm unit}} \def\tu{t_{\rm unit}} \def\Va{V_{\rm A}} \def\Av{\Alf velocity }
\def\sub{\subsection} 
\def\revise{}

\begin{abstract} 
Propagation of fast-mode magnetohydrodynamic (MHD) waves in the interstellar space is simulated, and focusing MHD wave (FMW) model is proposed for triggered star formation (SF). Waves from an SF region are trapped by nearby molecular clouds and converge onto their focal points, causing implosive compression. Even an isolated cloud suffers from long-distance invasion of waves from remote sources. Echoing SF occurs inside a cloud as well as between clouds. {\revise Repetitive refocusing in a filamentary cloud suggests spatial periodicity in SF sites along the filament.} The model is applied to the SF regions M16 and M17, where MHD waves produced by M16 are shown to converge onto the focal point of nearby GMC and trigger the SF in M17.
\end{abstract}

\section{Introduction} 
Triggering mechanisms of star formation (SF) in molecular clouds (MC) can be categorized into the following four types.
(A) Self-gravitational collapse of density fluctuation by the Jeans instability, which is the simplest mechanism occurring stochastically in a time scale as short as  $t\sim 1/\sqrt{G\rho}\sim 0.1-0.2$ My in MC cores with density $\rho \sim 10^5-10^6$ \Htwocc. However, in the majority of MCs with lower densities, the Jeans time is much longer, $\sim 10$ My. In such clouds, forced compression is proposed as a trigger, which includes
(B) the sequential SF, where the MC is compressed by expanding HII gas around an OB cluster (Elmegreen and Lada 1977), and 
(C) cloud-cloud collision, where the gas is compressed by encounter with other cloud (Scoville et al. 1986; McKee and Ostriker 2007).  
In case of an isolated cloud,
 (D) ambient interstellar disturbances are suggested to be effective trigger, which grow on the cloud's surface to cause implosive compression (Woodward   1978; Le{\~a}o et al. 2009; Boss et al. 2012).  

In mechanism (D), kinetic energy released at SF sites or supernovae is transported to a remote molecular cloud by sound, \Alf, and fast-mode (compression mode) MHD waves. The sound wave is too slow to travel for long distance, and \Alf wave cannot compress the gas. Therefore, among the three, only the fast-mode MHD wave (hereafter, MHD wave) is the possible conveyer of energy in order to trigger the SF. In this paper, a focusing MHD wave (FMW) mechanism is proposed based on simulation of the propagation of the waves in the interstellar space.

\section{Method } 

\subsection{Fast-mode MHD waves} 
The basic equations of motion, or the Eikonal equations, to trace the fast-mode MHD waves of small amplitude were obtained in order to study the Morton waves in the solar corona (Uchida 1970, 1974). The method has been applied to the Galactic Center explosion (Sofue 1977) and old supernova remnants (Sofue 1978). 
Given the distribution of \Av and initial direction of the wave vector, the equations can be numerically integrated to trace the ray path as a function of the time. As to the detail for solving the equations, see the above papers.  

 \subsection{Magnetic field and \Alf velocity} 
Zeeman effect observations of HI and OH absorption lines (Crutcher et al. 2010) show that line-of-sight strengths of magnetic field are $B_z\sim 1-10$ \muG in MCs with density $n_{\rm H}\sim 10-10^4$ \Hcc, and  $\sim 10^2-10^3$ \muG in high-density cores of $10^6 -10^7$ \Hcc. Far-infrared observations of dust polarization (Planck collaboration 2016) reveal that magnetic lines of force penetrate the clouds without particular change of flux. In larger scale, magnetic field in the local Galactic disc within a few kpc is shown to be about constant at strength of $B\sim 5-10$ \muG (Sofue et al. 2019). 

\begin{figure} 
\begin{center}  
\includegraphics[width=8cm]{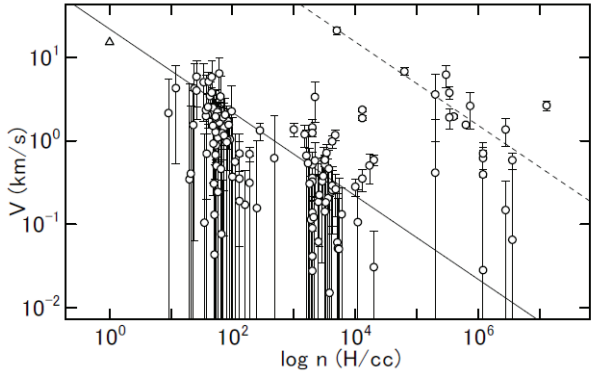}   
\end{center}
\caption{\Alf velocity plotted against H density of molecular clouds calculated for $B=\sqrt{2}B_z$ and $\nH$ from the observational data presented by Crutcher et al. (2010). Triangle indicates the local galactic value. The solid line represent $V=21.9\ (n_{\rm H}/1\ {\rm H\ cm^{-3}})^{-1/2}$ \kms for constant magnetic strengths normalized at $B=10$ \muG and $\nH=1$ \Hcc. There seems an upper branch fitted by the same relation but 70 times faster as represented by the dashed line.  }
\label{Va} 
	\end{figure}

Figure \ref{Va} shows \Alf velocities in MCs calculated for observed values of $B=\sqrt{2}B_z$ plotted against the gas density using the data from Crutcher et al. (2010), where $\sqrt{2}$ is a correction from line-of-sight to total strength. The triangle denotes the local interstellar value (Sofue et al. 2019). There appear two sequences in the plot. The main branch is composed of MCs with density less than $\sim 10^5$ \Hcc and is approximately fitted by a straight line for a constant magnetic field with $B =B_0=10$ \muG expressed as
$
V\sim 22 (n_{\rm H}/1{\rm cm^{-3}})^{-1/2}\ {\rm  km\  s^{-1}}.
\label{Alf}
$
In the simulation this relation is adopted for MCs and galactic disc, except for a high-density core. The sub-branch for high-density cores also follows the same relation as shown by the dashed line representing \Alf velocity 70 times that for the major branch.  The large displacement of the sub-branch from the main may manifest the frozen-in property of magnetic field from MCs to dense cores.          

 \subsection{ Distributions of gas density and \Alf velocity} 

 The gas density of MC is represented by a Gaussian profile as $\rho=\rho_0 \exp(-(s/a)^2)$, where $a$ is the scale radius and $s$ is the distance from the cloud center. The galactic disc is represented by a plane-parallel layer with density profile $\rho=\rho_0 \sech(z/h)$, where $z$ is the height from the galactic plane and $h$ is the scale height.
The distribution of interstellar gas and clouds is represented as 
\be
\rho=\rho_0+\rho_d\ \sech \(z\/h \)+\Sigma_{i=1}^n \rho_i \
e^{-\Sigma_{k=1}^3\(x_k-x_{k,i}\/a_i\)^2},
\ee
where $\rho_0$ is a constant of the background gas density, $\rho_d$ is the density of disc gas at the galactic plane, and $\rho_i$, $x_{k,i}$, $a_i$ ($k=1,2,3$ for $x,y,z$) are the central density, position, and scale radius of the $i$-th cloud, respectively. 

The trajectory of the ray path of an MHD wave depends only on the functional form of the \Av, but not on its strength $V_0$. So, the following calculations are made in the non-dimensional scheme with the unit of time measured in scale unit$1/V_0$.

 \subsection{ Units and timescale}  

The equations are non-dimensionalized. The real quantities are obtained  using the units of length $A$, time $\tu$, and velocity  $\Vu$ through
$
\varpi=r A,\  \xi=Ax,\ \eta=Ay,\ \zeta=Az,\ \alpha=Aa,\ \tau=\tu t,
$
which represent the radius, Cartesian coordinates, cloud's size, and time, respectively. The unit of velocity is given by the \Av of the background ISM,
$
\Vu=\Bu/\sqrt{4\pi \rhou}.
$
In this paper the following values are adopted:
$\rhou=1$ \Hcc, $\Bu=10 \mu$G, $A=100$ pc, leading to 
$\Vu=\Bu/\sqrt{4 \pi \rhou}=21.892$ \kms and $\tu=4.487$ My.  
The results are valid for different density and magnetic field units, where the time unit may be changed to $4.487 \times \sqrt{\rho/\rhou}/(B/\Bu)$ My. Namely, the denser is the cloud and/or the weaker is the magnetic field, the longer is the time scale, corresponding to slower propagation with smaller \Alf velocity.

\section{2D Calculations}

 \subsection{Implosion onto a focal point in a cloud} 
\begin{figure*} 
\begin{center}  
\includegraphics[width=14cm]{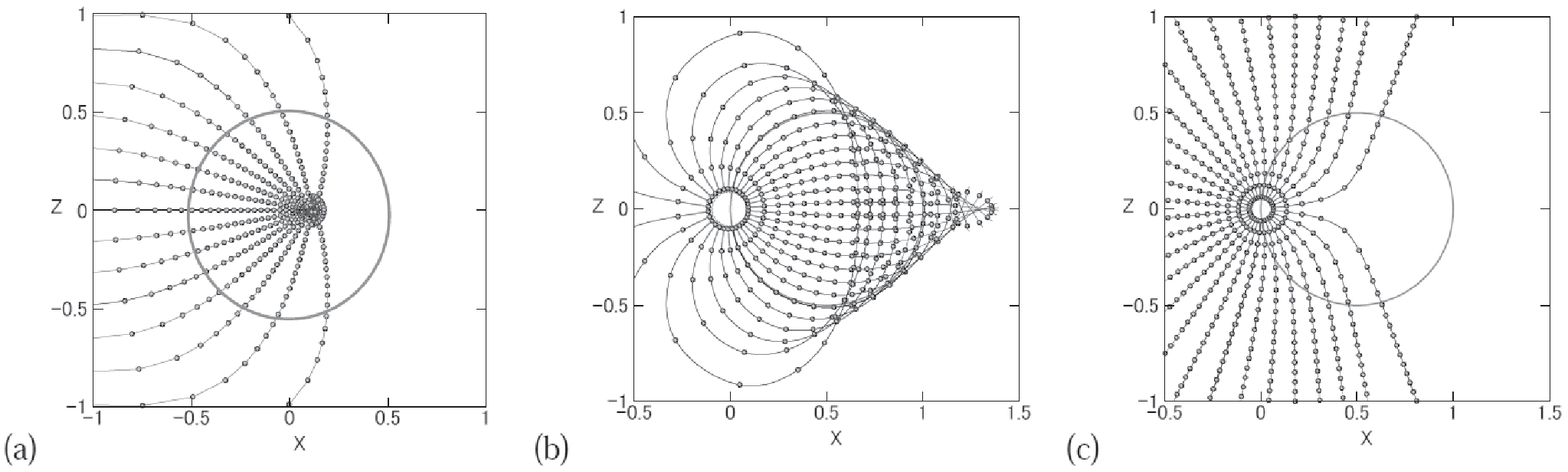}  
\end{center}
\caption{(a) Plane wave focusing on a Gaussian cloud at $(x,y)=(0,0)$ with scale radius $a=0.5$ (big circle) plotted every $\delta t=0.1$ for density distribution 
$\rho=0.1+10 \exp(-(x^2+z^2)/0.5^2)$.
(b) Spherical wave from the edge of a cloud at $(x,y)=(0.5, 0)$ focusing on the opposite side for 
$ \rho=0.1+10 \exp(-((x-0.5)^2+z^2)/0.5^2)$  
{\revise (c) Same, but for a void of molecular gas with $\rho=1.0-0.95\exp(-((x-0.5)^2+z^2)/0.5^2)$ every $\delta t=0.05$.  }}
\label{mhd_1cloud} 
	\end{figure*}

Propagation of a fast-mode MHD wave encountering an isolated molecular cloud is traced by solving the Eikonal equations. The result is shown in figure \ref{mhd_1cloud}, where a plane and spherical waves encounter a cloud with Gaussian density profile. The waves are refracted by the cloud, in which the propagation velocity decreases toward the cloud center. The cloud properties such as the density profile and used parameters are presented in individual figure captions. The clouds play a role of convex lens for the MHD rays.

The focal length for a plane wave is as short as $\sim 0.1a$,  where $a$ is the scale radius of the cloud. The wave converges onto a focal point at an efficiency as high as $\sim 99$\% of the wave within an impact parameter of $\sim 2a$, and gathered into a focal sphere of radius $\sim 0.02a$. This means that a Gaussian cloud is a good convex lens with small optical aberration.

        \begin{figure} 
\begin{center}  
\includegraphics[width=8cm]{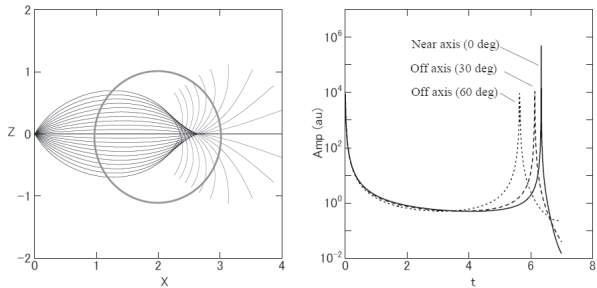}  
\end{center}
\caption{(left) Near-axis spherical wave focusing on a cloud of radius $a=1$ centered on $x=2$ for 
$ \rho=0.1+10 \exp(-((x-2)^2+z^2)/0.5^2)$. 
(right) Relative amplitudes in arbitrary unit plotted against elapsed time since the point injection for near- and off-axis waves radiated at $\theta=0\deg, 30\deg, 60 \deg$. }
\label{flux} 
	\end{figure}  
  
The focusing occurs also for a spherical wave originating at a point close to the cloud. A case for a wave source at the half-density radius is shown in figure \ref{mhd_1cloud}(b). Although the aberration is stronger than that for a plane wave, the focusing efficiency is still as good as $\sim 50$\% onto the focal sphere of radius $\sim 0.2a$ at $\sim 1.5a$ from the cloud center on the opposite side. It is shown that a molecular cloud with Gaussian profile is a good collector of MHD disturbances, which gathers a significant fraction, $\sim >50$\%, of the waves encountering the cloud within the impact parameter of a few cloud radii. In other words, any isolated molecular cloud is expected to experience implosive compression by invasion of disturbances from distant SF sites.

The amplitude of the focusing wave increases toward the focal point. Figure \ref{flux} shows time variation of the amplitudes of the near- and off-axis waves at $\theta=0, 30\deg, 60\deg$ in arbitrary unit against the elapsed time since the point injection at the origin for a cloud centered on $x=2$ with $a=1$. The flux (amplitude) increases drastically toward the focus.

 \subsection{  Focusing efficiency for different type of clouds} 
Once MHD waves encounter a molecular cloud, they are converged toward the focal point. However, density profiles in molecular clouds may not be so simple. So, several different types of density profiles are examined, adopting the following models.
\be
\rho=\rho_0+\rho_{\rm c} f(x,z),
\ee
where $\rho_0=0.1$, $\rho_{\rm c}=1$, and the function $f$ is given by\\
(i) $f=\exp[-((x-a)^2+z^2)/a^2] $,\\
(ii) $f=1/(\epsilon+((x-a)^2+z^2)/a^2)$, \\
(iii) $f=\exp[-((x-a)^4+z^4)/a^4]$, and \\
(iii) $f=1/(\epsilon+((x-a)^4+z^4)/a^4)$.\\
The \Alf velocity is calculated for a constant magnetic field of $B=B_0=1$ in unit defined in the previous subsection.

	\begin{figure} 
\begin{center}   
\includegraphics[width=8cm]{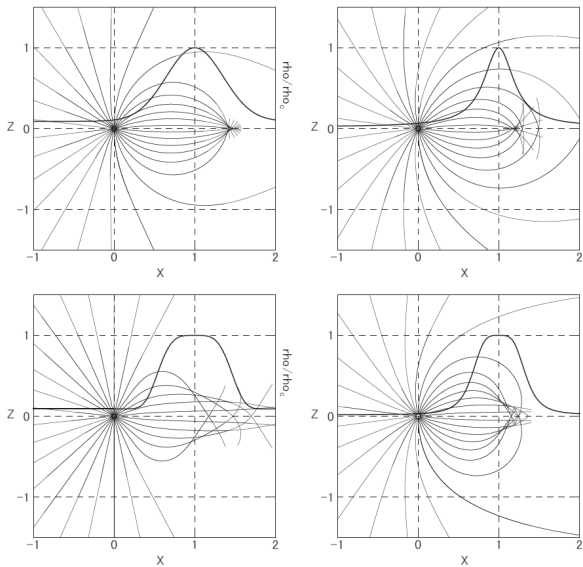}  
\end{center}
\caption{MHD ray refraction by a cloud centered on $(x,z)=(1,0)$ and scale radius $a=0.5$ for (tl) Gaussian density profile; (tr) 2nd-order politropic profile mimicking an isothermal gas sphere in gravitational equilibrium, (bl) 4th order Gaussian profile mimicking plateued density cloud with sharp edge, and (br) 4th politropic profile, mimicking a plateaued-density cloud with a mild edge. Circles indicate the position and scale radius of clouds, and thick lines show density profiles along the $x$ axis normalized to unity at the cloud center. }
\label{NoGauss} 
	\end{figure} 
        
Case (i) represents a Gaussian cloud as used in the previous sections; (ii) represents a semi-isothermal gas sphere in gravitational equilibrium with the density proportional to the inverse square of the radius, or the second politropic index; (iii) represents a plateaued cloud with a sharper edge, mimicking a flattened cloud; and (iv) a flattened cloud of fourth politropic index. 
Figure \ref{NoGauss} shows calculated MHD ray paths released at $(x,z)=(0,0)$ which propagate through clouds of radius $a=0.5$ located at $(x,z)=(1,0)$. The density parameters are taken to be $\rho_0=0.1$, $\rho_{\rm c}=1.0$, and $\epsilon=0.2$. Gas density in each cloud normalized to the value at the cloud centers is shown by thick line.

The MHD ray paths in cases (i) and (ii) show similar behavior, both tightly focusing on the focal points. Accordingly, the wave amplitude increases steeply toward the focuses as seen in figure \ref{flux}. Although both have similar lensing properties, Gaussian cloud has a wider impact area of focusing than isothermal cloud, and hence higher efficiency in the feedback for the triggering SF at the focal point.
On the other hand, cases (iii) and (iv) show milder convergence, and have no definite focal point, although convergence into a narrow region on the axis occurs as well.

 \subsection{ Oblique cloud and filament} 
\begin{figure*} 
\begin{center}     
\includegraphics[width=14cm]{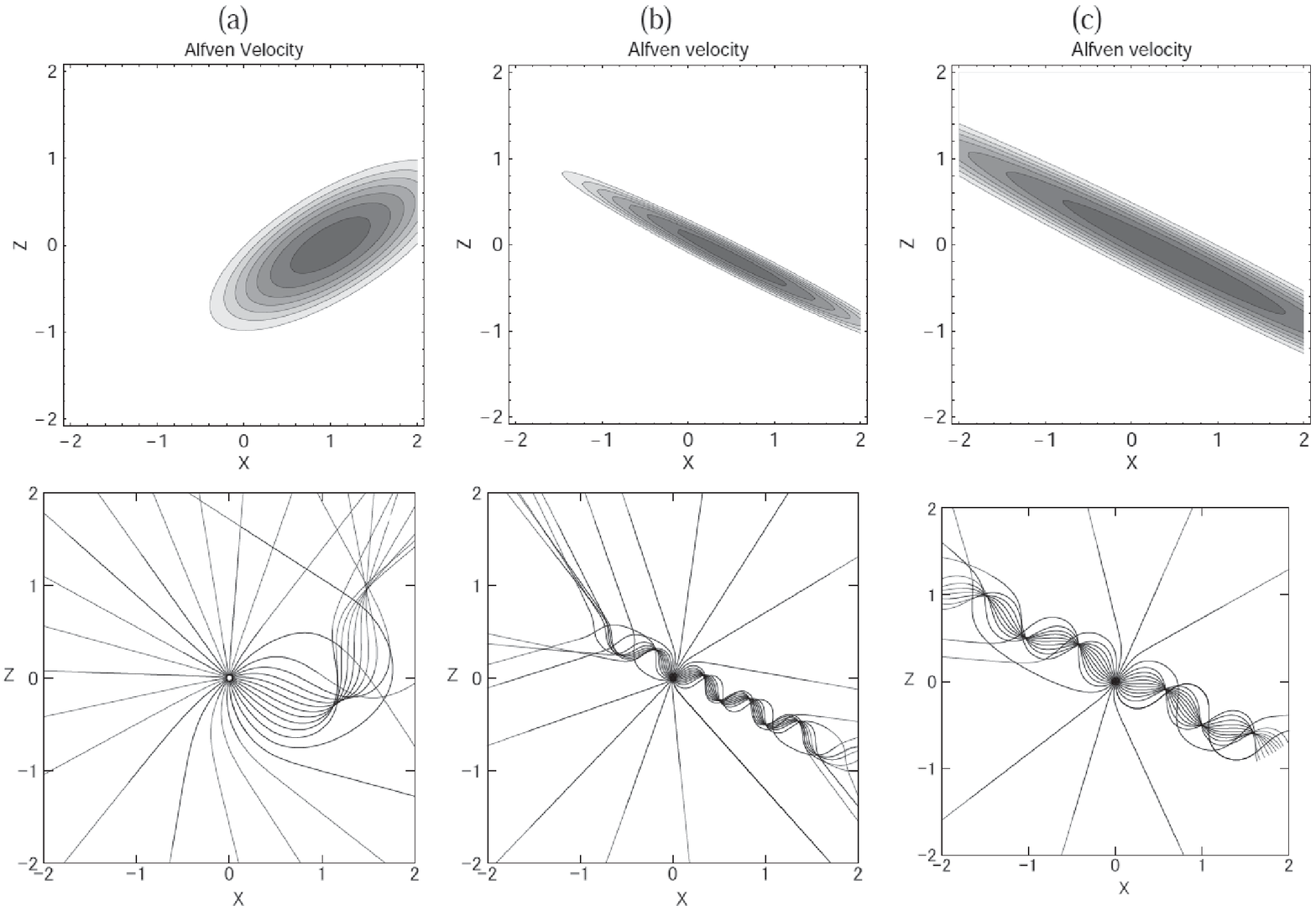} 
\end{center}
\caption{(a) Double focusing of MHD waves refracted by an oblique elliptical cloud having elliptical Gaussian profile with the major axis inclined by 45$\deg$ from the $x$ axis, expressed by
$\rho= 0.1 + 1.0 \exp(-((z^2+(x-z-1)^2)/0.5^2)) $.   
(b)MHD waves guided within a filamentary cloud repeating reflection at the edges as a wave guide. The cloud's density profile is given by 
$\rho= 0.1 +1.0 \exp(-(((0.5 x+z-0.1)/0.1)^2+((x-0.5)/1.0)^2)) $.
{\revise (c) Same as (b), but the size of filament is doubled. Waves from the near axis origin is reflected and guided along the filament axis, focusing many times around the axis.}
 }
\label{oblique} 
\label{filament} 
	\end{figure*}

It often happens that clouds are elongated. In order to qualitatively discuss the effect of non-sphericity, an elliptical cloud with Gaussian density profiles in the major and minor axes with axial ratio of 2 was examined. The cloud is located obliquely at $(x,z)=(1,0)$ at an inclination of $45\deg$ from the $x$ axis. Figure \ref{oblique} shows the result. 

The rays are first converged to a focal point on the opposite side with large aberration due to the off-set ray paths from the axes. They are then refracted and reflected by the far-side edge, and are again converged onto another focal point on the same side of the source. The second focusing occurs at a much higher efficiency, where the "aberration is well corrected". The result indicates that triggered SF can happen at multiple places in the same cloud.  

Another case for a filamentary cloud, where the wave is emitted at an off-center edge, is shown in figure \ref{filament}. The wave is strongly reflected by the cloud's surface working as a wave duct, and is guided along the major axis, until it escapes from the axis ends. 
{\revise If the wave source is located nearer to the cloud axis, the focusing occurs at higher efficiency. Figure \ref{filament}(c) shows a similar case in a two times bigger filamet, where the waves repeat refocusing at higher efficiency, having periodicity with interval of about cloud's width.}

 \subsection{  Diffraction by frozen-in magnetic core} 
The assumption of constant magnetic field is obviously too simplified, particularly in dense cores, where the \Alf velocity attains higher values as observed in figure \ref{Va}. This means that the focusing on the densest core is weakened, so that the trigger of SF is rather suppressed in such a high-magnetic core. 

\begin{figure} 
\begin{center}  
\includegraphics[width=8cm]{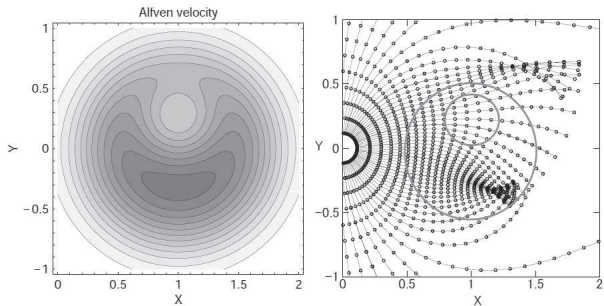}  
\end{center}
\caption{
Distribution of the \Alf velocity and MHD ray paths exhibiting off-set focusing. The main cloud has density and $V$ profiles as
$\rho_1=0.1+ \exp(-((x-1)/0.5)^2+(x/0.5)^2)$ and $V_1=1/\sqrt{\rho_1}$.
The cloud is superposed by an off-center core with 
$\rho_2=100 \exp(-((x-1)/0.25)^2+((y-0.2)/0.25)^2)$ and frozen-in field,
having $V_2=\rho_2^{1/6}$. 
The net \Alf velocity is calculated by
$V=(V_1^2+V_2^2)^{1/2}$.
}
\label{frozenincore} 
	\end{figure}

In order to model such a case, ray paths are calculated through a cloud with two components as shown in figure \ref{frozenincore}. The main cloud has the same property as those in the previous sections, on which an off-set core with small diameter is superposed by frozen-in magnetic field having \Alf velocity proportional to $\rho^{1/6}$ of the core density. 
The net \Alf velocity is calculated by $V=\sqrt{V_{\rm cloud}^2+V_{\rm core}^2}$. The ray paths are diffracted by the high-$V$ core, but instead they focus on an off-axis focal point in the main cloud.

This simple simulation suggests that the wave focusing takes place somewhere in the cloud, even if the cloud has structures with different magnetic properties. Also, the focal point is not necessarily coincident with the highest-density core.

 \subsection{  Focusing in the disc and clouds}      
\begin{figure} 
\begin{center} 
\includegraphics[width=8cm]{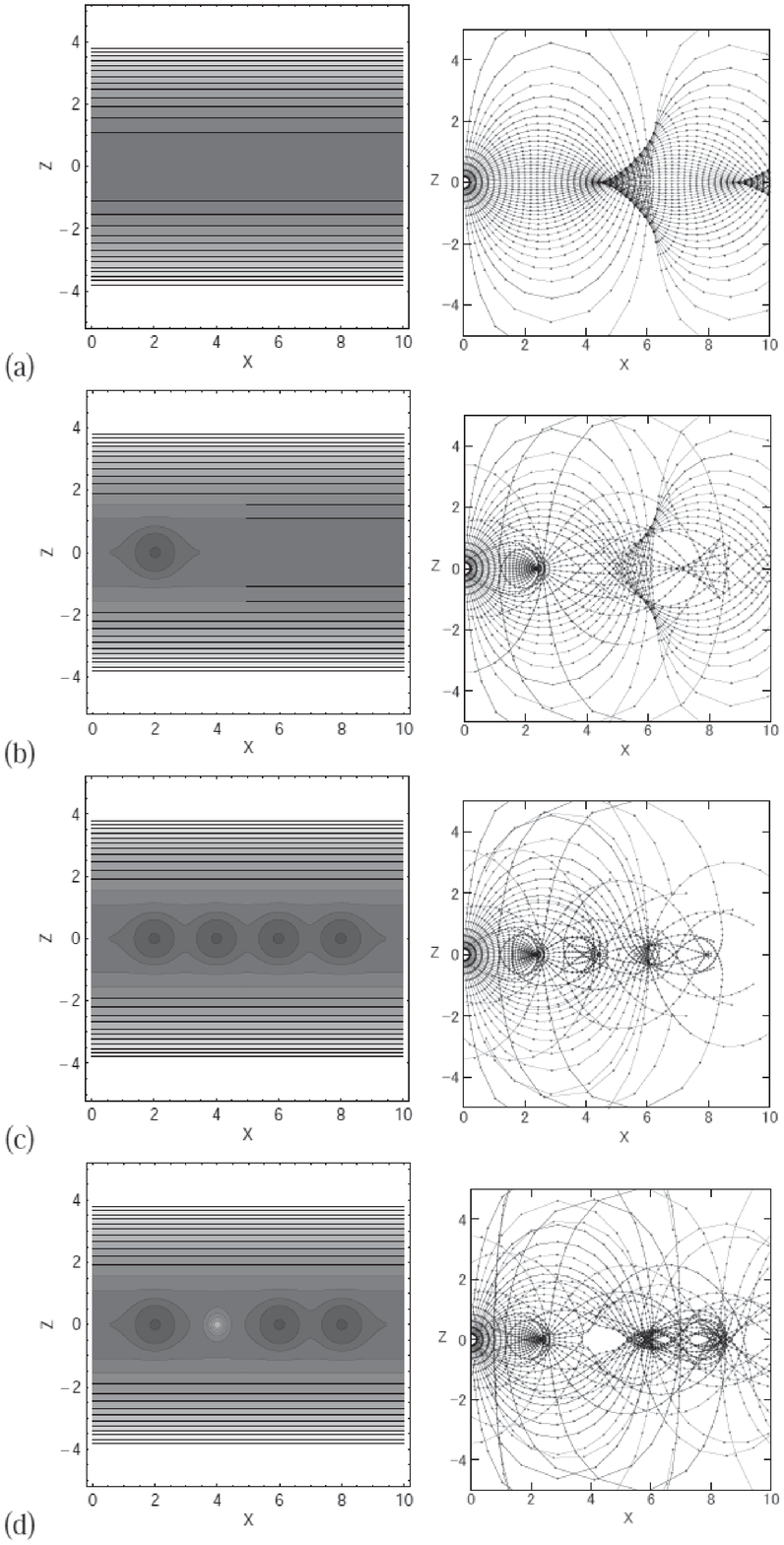} 
\end{center}
\caption{(a) MHD wave from a point, propagating through a stratified gas layer with sech density profile as
$\rho=0.1 +\sech(z/1.0) $. 
The wave front is plotted every $\delta t=0.1$ by the small circles.
(b) Same, but encountering a cloud at $x=2$  with
$\rho=0.1+1.0 \sech (z/1.0 ) +5.0 \exp(-(z^2+(x-2)^2)/0.5^2)$.
(c) Same, but with clouds at $x=2$, 4, 6 and 8.
{\revise (d) Same as (c), but the second cloud was replaced by a void.} 
}
\label{mhdDisk} 

\begin{center}  
\label{mhdMeshSech}  
\includegraphics[width=8cm]{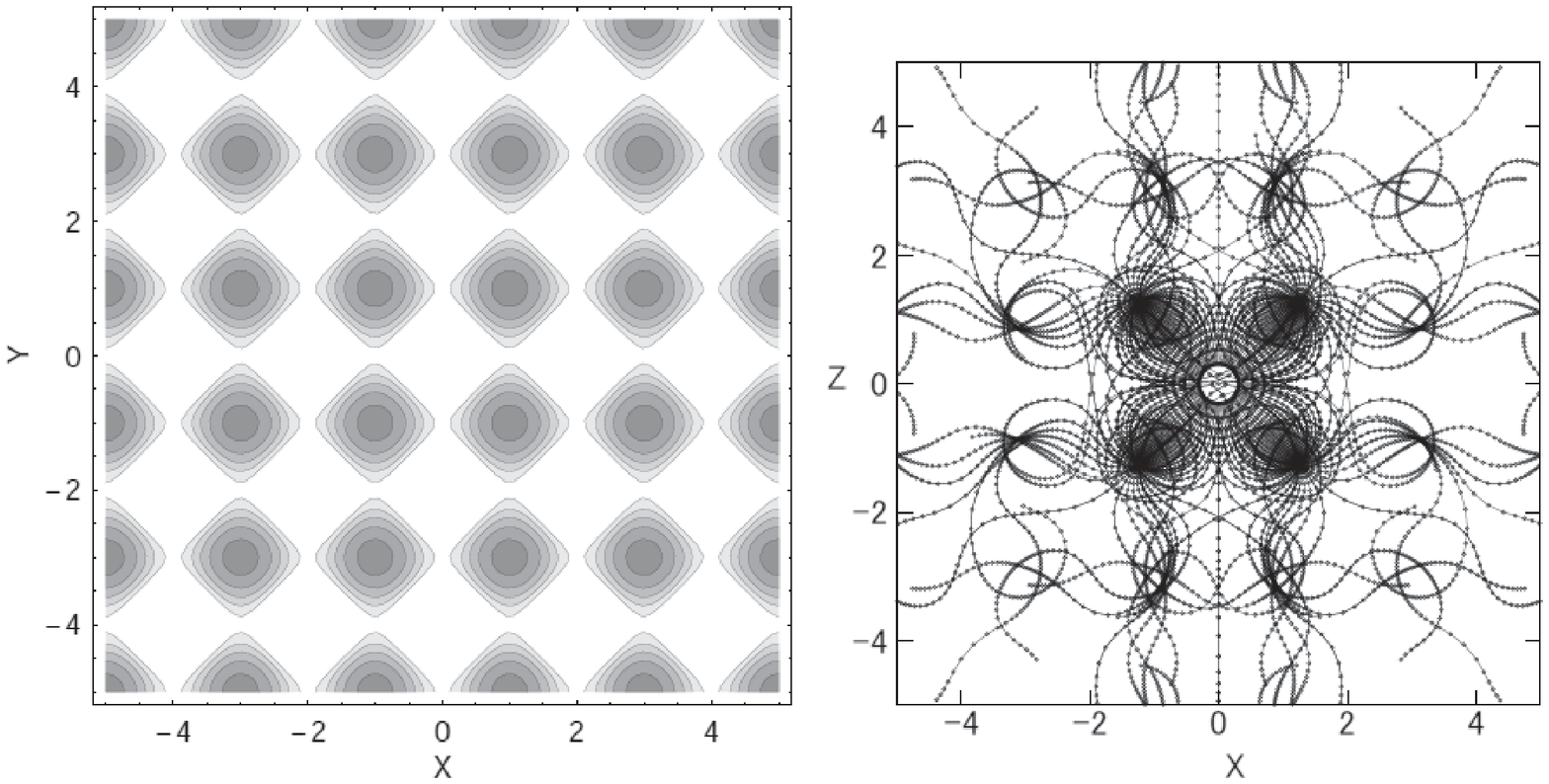}    
\end{center}
\caption{Gas clouds in the galactic plane on squared mesh points as
$\rho=0.1+5.0 \sech(z/1.0)(\exp(-2.(\sin(\pi x/2)^2+\sin(\pi  z/2)^2))$, and MHD wave fronts emitted between the clouds at every $\delta t=0.1$.  
} 
\label{mhd_mesh} 
	\end{figure}

Larger scale propagation of MHD waves in the Galactic disc is also subject to focusing onto MCs. Figure \ref{mhdDisk}(a) shows a case of a spherical wave originating in the galactic plane. The wave is reflected by the rapidly increasing \Alf velocity toward the halo, and converges to a focal ring of radius $x\sim 4.4h$. After passing the focus, the wave expands again, reflected, and converges to the next focus at $x\sim 8.5h$. Thus, the galactic disc is a good container of the interstellar disturbances, where a released MHD wave is confined within the disc at a high efficiency, and periodically converges to focal rings of radii $\sim 4.4 n h$ with $n=1, 2,$ ...
        
If there exists a cloud in the disc as shown in figure \ref{mhdDisk}(b), the rays are focused on its focal point determined by the cloud's radius and the distance from the wave source. After passing the focus, the rays further repeat expansion and convergence.

If there are a number of clouds, as in figure \ref{mhdDisk}(c), the rays once focused to the first cloud repeatedly converge to the other clouds focal points.
Thus, molecular clouds in the galactic disc suffer from imploding injection of MHD waves that were generated remote sources.

Figure \ref{mhd_mesh} shows a case when many clouds are distributed in the galactic plane, and a spherical MHD wave is released between the clouds. The MHD rays are converged to neighboring clouds first, then repeat expansion and convergence onto more distant clouds. The waves are well confined into the clouds, whereas the inter-cloud regions play a role as a fast transmitter of the waves to the nearby clouds. 

From the simulations, it may be concluded that interstellar fast-mode MHD waves originating at a SF site are well confined within the galactic disc, and converge onto neighboring clouds at high efficiency. Even waves escaped from a cloud inevitably trapped and converge to other clouds. The waves converged on focal points inside the clouds will compress the gas to trigger SF. In other words, one single SF activity or a supernova explosion causes to trigger SF in many other remote clouds in the disc.

{\revise
\subsection{Voids}
The interstellar space is often observed to have cavity of ISM, or void, filled with high temperature and low density gas supposed to be formed by explosive events such as supernovae and/or stellar winds from massive stars. Although expelled at the same time, magnetic field is considered to be less affected, remaining or recovering soon, for its tensional property. This causes higher \Av  in the voids, where the MHD wave is expected to be reflected.
In order to examine how such voids affect the MHD wave propagation, we traced wave propagation through ISM where a cloud is replaced by a void defined by low-\Av region having the same extent.  

Figure \ref{mhd_1cloud}(c) shows a case, where a spherical MHD wave encounters with a nearby void. The waves are reflected and scattered, and the void acts as a concave lens. Figure \ref{mhdDisk}(d) shows a case when a nearby cloud in a disc is replaced with a void. The waves are reflected and passed through to the next cloud, where the convergence occurs at higher efficiency than in figure \ref{mhdDisk}(c). Furthermore, the void accelerate the propagation because of the faster transmission due to higher \Av.
}
\section{3D Calculations}
In this section 3D calculations are performed in order to display the propagation of the MHD wave fronts in a more realistic condition, where the line of sight is not necessarily parallel to the wave front. For this purpose a thousand of rays are initially distributed randomly on a small sphere with radial direction. 

 \subsection{  Single cloud}

\begin{figure} 
\begin{center}  
\includegraphics[width=8cm]{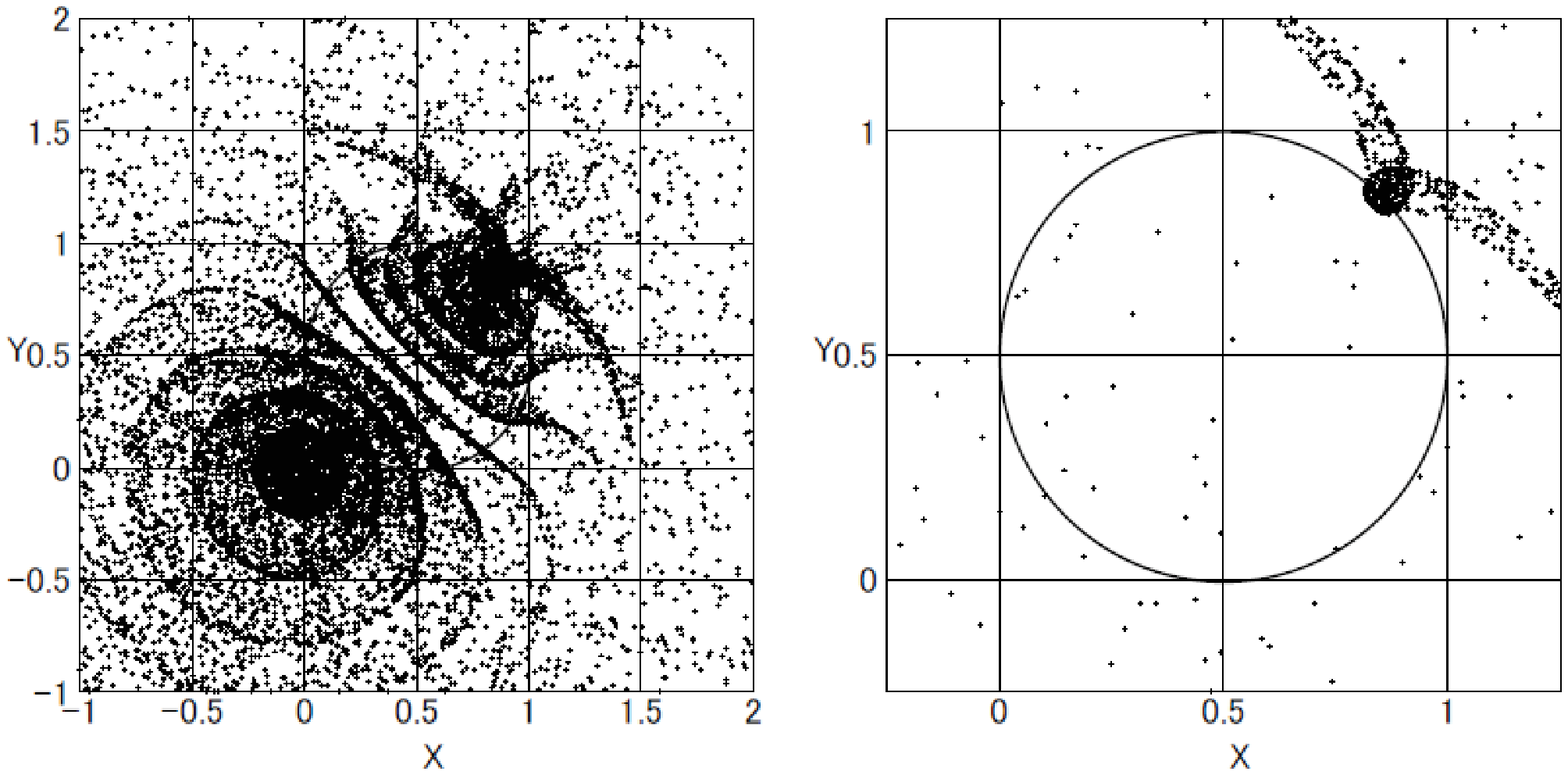}   
\end{center}
\caption{3D self focusing of MHD waves on the opposite side of the cloud projected on the X,Y plane from $t=0$ to 3.3 plotted every $\delta t=0.3$, and a close up of snap shot at $t=3.2$. The cloud is centered on (0.5, 0.5, 0) with radius 0.5 with density
$\rho=1+10\exp(-(((x-0.5)/0.5)^2+((y-0.5)/0.5)^2+(z/0.5)^2)) $ }
\label{3d_1cloud} 

\begin{center}  
\includegraphics[width=8cm]{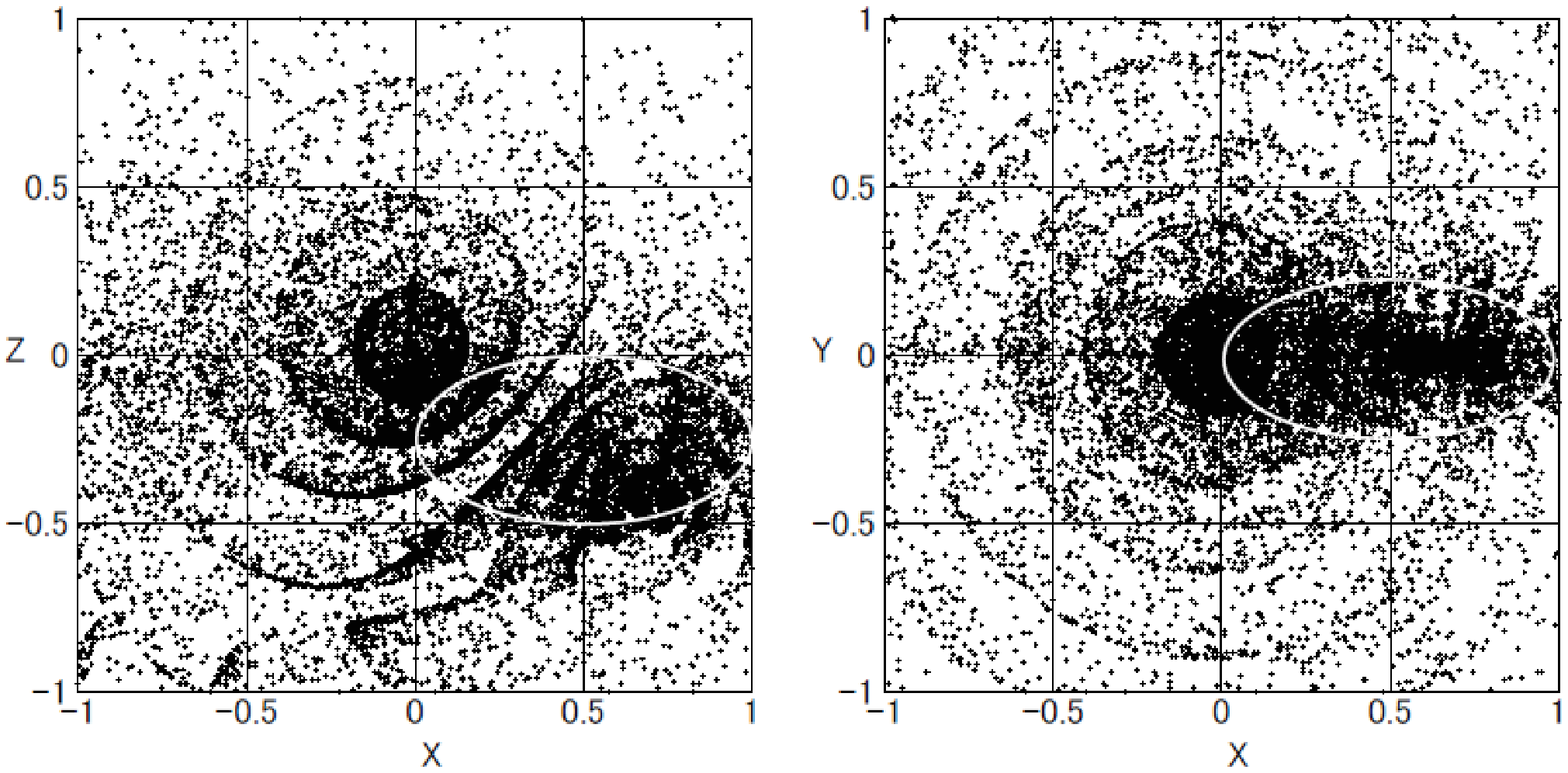}   
\end{center}
\caption{3D oblique self focusing in an elongated cloud with $\rho=1+10\exp(-(((x-0.5)/0.5)^2+(y/0.5)^2+((z-0.5)/0.25)^2)) $. Wave is plotted every $\delta t=0.3$. }
\label{3d_self} 
	\end{figure}

Figure \ref{3d_1cloud} shows a result for a case when a Gaussian cloud is located in the galactic plane at $(x,y,z)=(0.5,0.5,0)$ with a radius of 0.5. The propagation of the rays are almost the same as calculated in the 2D simulation, while the deformation of the spherical front can be more visually traced. The focusing efficiency is so high that the wave radiated in a $90\deg$ (in diameter) cone facing the cloud converges onto the focus.

Figure \ref{3d_self} shows a case when the cloud is elongated in the $y$ direction, and is located off axis from the ray ejection at the origin, representing an oblique encounter with non-spherical cloud. Even in this case, the considerable fraction of the wave front is converged to the focus on the opposite side of the cloud at high efficiency.
  
 \subsection{  Multiple clouds and voids} 

\begin{figure} 
\begin{center}  
\includegraphics[width=8cm]{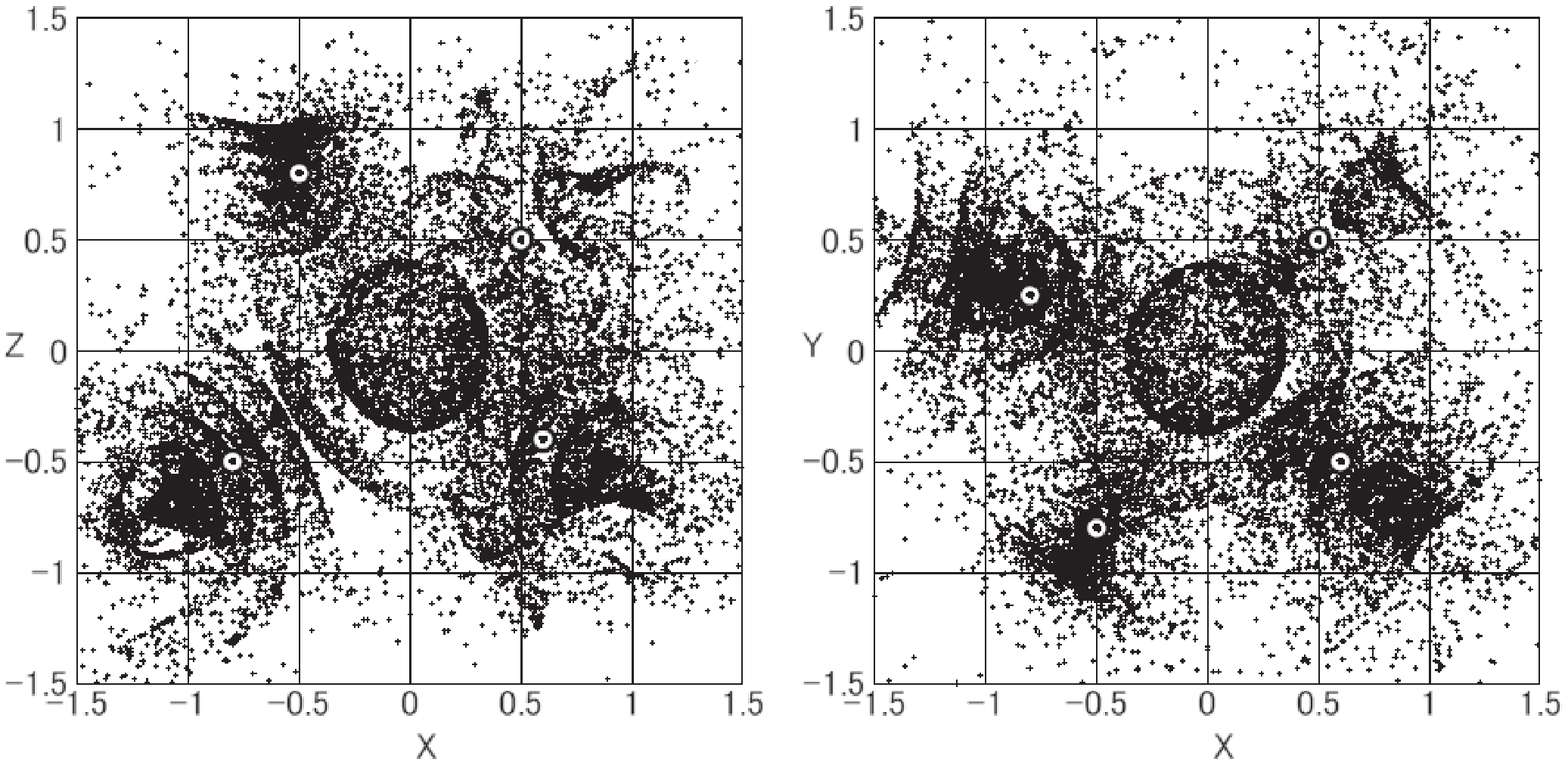}     
\end{center}
\caption{ 3D inter-cloud convergence onto multiple clouds projected on (X,Z) and (X,Y) planes. The density distribution is given by
$\rho=1.0 \sech(z/0.5)
+10\exp(-(((x-0.5)/0.5)^2+((y-0.5)/0.5)^2+((z-0.5)/0.5)^2)) 
+20\exp(-(((x-0.6)/0.5)^2+((y+0.5)/0.5)^2+((z+0.4)/0.5)^2))
+30\exp(-(((x+0.8)/0.5)^2+((y-0.25)/0.5)^2+((z+0.5)/0.5)^2))
+50\exp(-(((x+0.5)/0.5)^2+((y+0.8)/0.5)^2+((z-0.8)/0.5)^2)) $. 
Wave is plotted every $\delta t=0.7$. 
}
\label{3d_4cloud} 
\begin{center}   
\includegraphics[width=8cm]{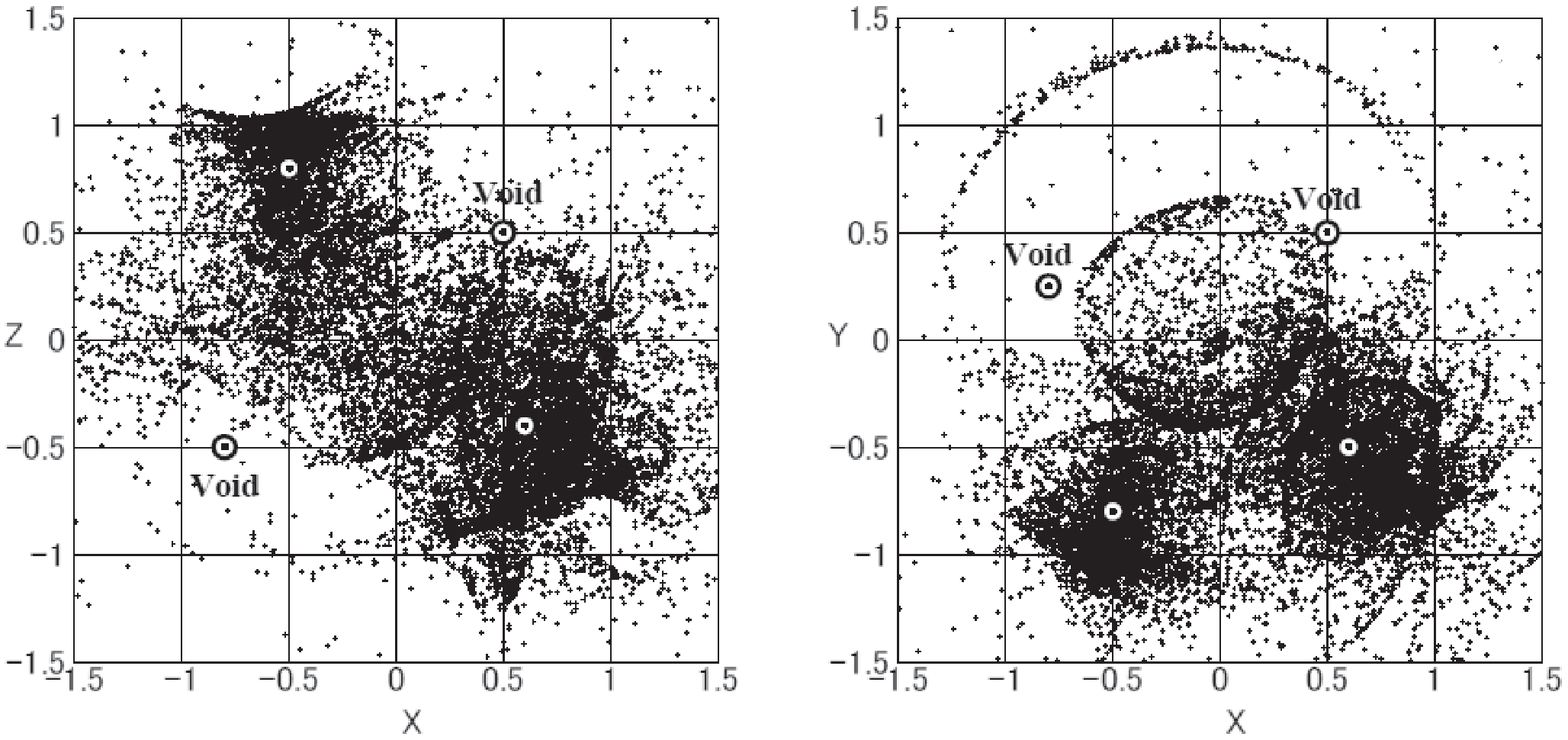}   
\end{center}
\caption{ {\revise  Same as figure \ref{3d_4cloud}, but two nearby clouds are replaced with voids of amplitude $-0.95$. Note the more efficient convergence on the other clouds.}}
\label{void_3d_4cloud} 
	\end{figure}

A case when several clouds are randomly distributed around the wave source is shown in figure \ref{3d_4cloud}. Here, four clouds of different sizes and densities are present around the wave source. The front expands and strongly deformed, and further diverges to four concave fronts facing the four clouds. Then, the waves are all absorbed by the clouds, and converge onto individual focal points. The 3rd panel of figure \ref{3d_4cloud} is a close up of one of the clouds, showing that each converging efficiency is as high as in the previous cases for single cloud. 

This means that almost all the waves radiated from the source is absorbed by these four clouds at an extremely high rate. Namely, all the ejected energy from the central source is shared by the four sources, and are converged onto the small volumes around the focal points.

Another important point is that the focal points appear on the opposite sides of the clouds with respect to the wave source. This is is an essentially different point from the other triggering mechanism of SF, in which the triggering takes place at the surface of the cloud facing the encountering shock wave or UV radiation. 

{\revise Figure \ref{void_3d_4cloud} shows a case where nearby two clouds are replaced by voids with higher \Av. The waves are reflected by the voids, and are more effectively converged to the other clouds. This simulation suggests that the triggering of SF by MHD focusing is more effective, when the inter-cloud space is filled with voids and cavities.
}

 \subsection{  Remote focusing} 

\begin{figure} 
\begin{center}  
\includegraphics[width=8cm]{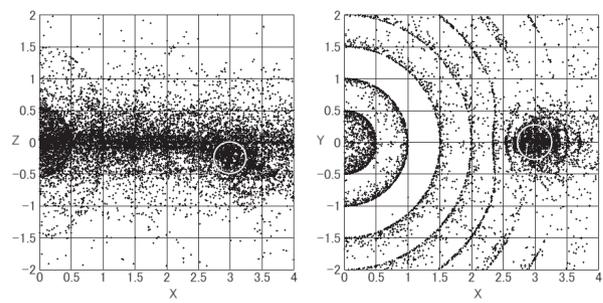}   
\end{center}
\caption{3D convergence to an isolated cloud after long-distance confined propagation through the sech disc with
$\rho=1.0 \sech(z/0.5) +20\exp(-(((x-3)/0.5)^2+(y/0.5)^2+((z-0.25)/0.5)^2)) $. Waves expands as a ring, not a sphere, because of th confinement by the sech disc. Wave is plotted every $\delta t=0.5$.}
\label{3d_long} 
	\end{figure}

Figure \ref{3d_long} shows a case where the wave front propagates through a sech disc of scale height $h=0.5$, and an isolated cloud of radius $a=0.5$ is put far remote from the source at $(x,y,z)=(3,0,-0.25)$.

Due to the high-efficiency reflection by the increasing \Alf velocity toward the halo, all the MHD wave front is tightly confined in the galactic disc. As the front expands and approaches the cloud, the wave front within a cone of $\delta \phi \times \delta \theta \sim 40\deg \times 180\deg$ (diameter) is converged to the focal point at a high efficiency.
Thus, even an isolated cloud in the disc, far from the disturbance source, suffers from an invasion of the wave into its focus at a significant rate. 
  
\section{M16 and M17}

Possible evidences for remote triggered SF have been obtained in the local HII regions M16 and M17: Based on radio continuum, HI and CO-line maps and ages of OB clusters, Sofue et al. (1986) suggested that SF in M17 may have been triggered by a shock wave driven by M16. {Comer{\'o}n and Torra} recently analyzed the ages and elocutionary sequences of young stellar objects in and around M16 and M17, and concluded that the two HII regions may have common origin, both triggered by an explosive events in a supposed older cluster in between the two HII regions.

	\begin{figure} 
\begin{center}   
\includegraphics[width=6cm]{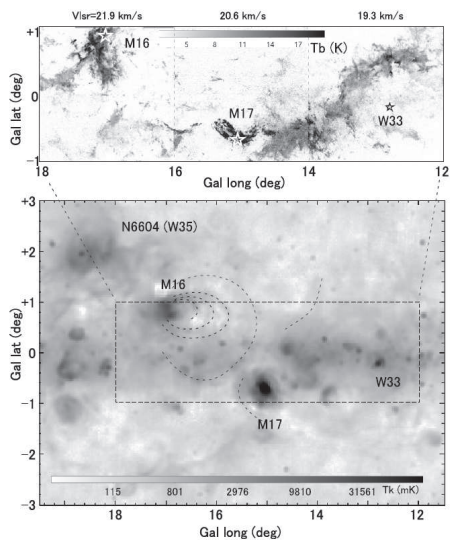} 
\end{center}
\caption{\co $\Tb$ map at $\vlsr \sim 20$ \kms around M16 and M17 molecular complex with a bar in K, and Bonn 11 cm radio contour map with grey scale in mK. Dashed lines are possible radio fronts, where the shells around M16 are those discussed in Sofue et al. (1985). }
\label{M16M17} 
	\end{figure} 

	\begin{figure} 
\begin{center}     
\includegraphics[width=8cm]{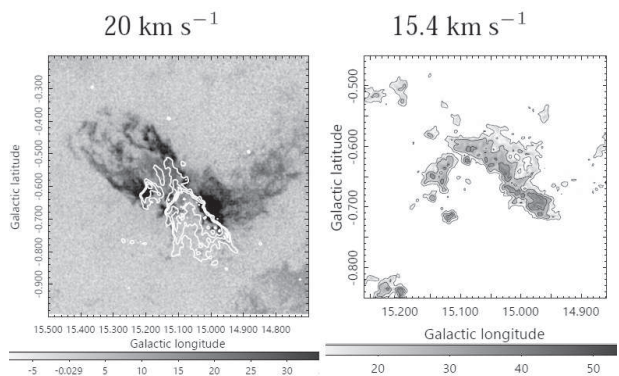} 
\end{center}
\caption{ 
\co-line brightness at 20 \kms around M17 overlaid by contours of radio continuum brightness at 90 cm, showing blistered HII region toward the south,
and close up of molecular half-shells at different centers at 15.4 \kms. }
\label{m17_coshells} 
	\end{figure} 
        
	\begin{figure*} 
\begin{center}   
\includegraphics[width=14cm]{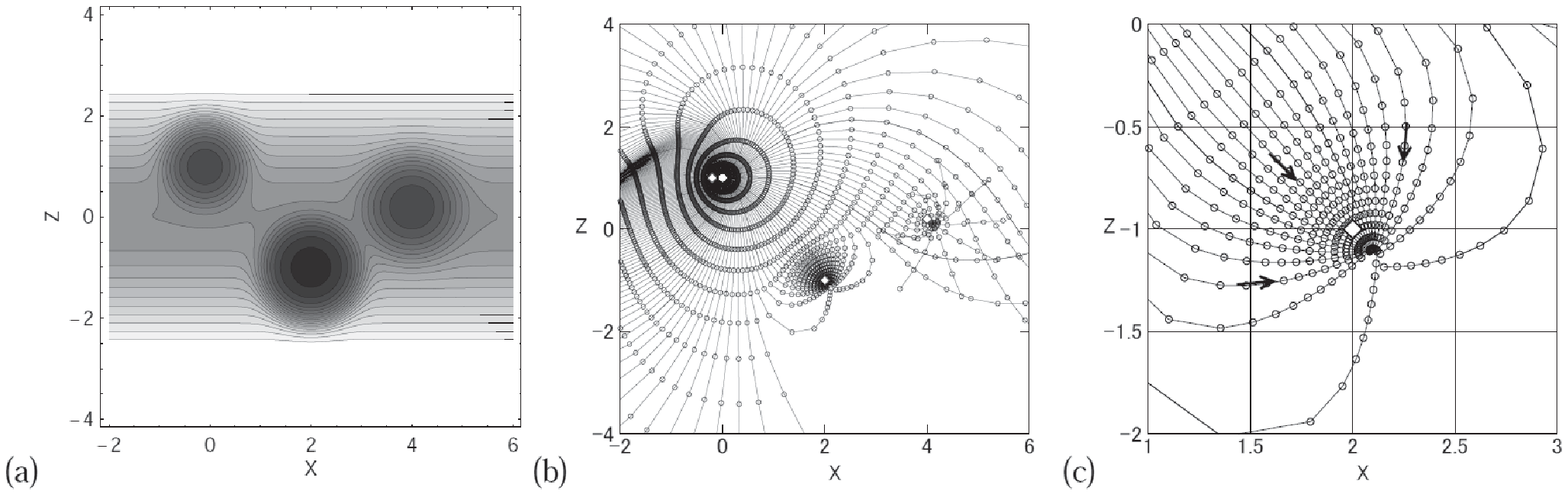} 
\end{center}
\caption{(a) 2D Model density distribution of molecular gas around M16 and M17.
 Clouds have peak densities 10, 50 and 10, respectively, and located in the sech galactic disc as
$\rho=0.1+1.0 \sech(z/1.0) 
+10.\exp(-(z^2+(x+0.2)^2)/0.5^2)
+50.\exp(-((z+2)^2+(x-2)^2)/0.5^2)
+10.\exp(-((z+0.8)^2+(x-4.0)^2)/0.3^2)$.
 Waves are plotted every $\delta t=0.4$.
(b) MHD simulation of remote SF triggering, mimicking M16 and M17. (c) Implosion of the wave on the focal point of the M17 cloud on the opposite side from M16.}
\label{mhdM16} 
        
\begin{center}   
\includegraphics[width=14cm]{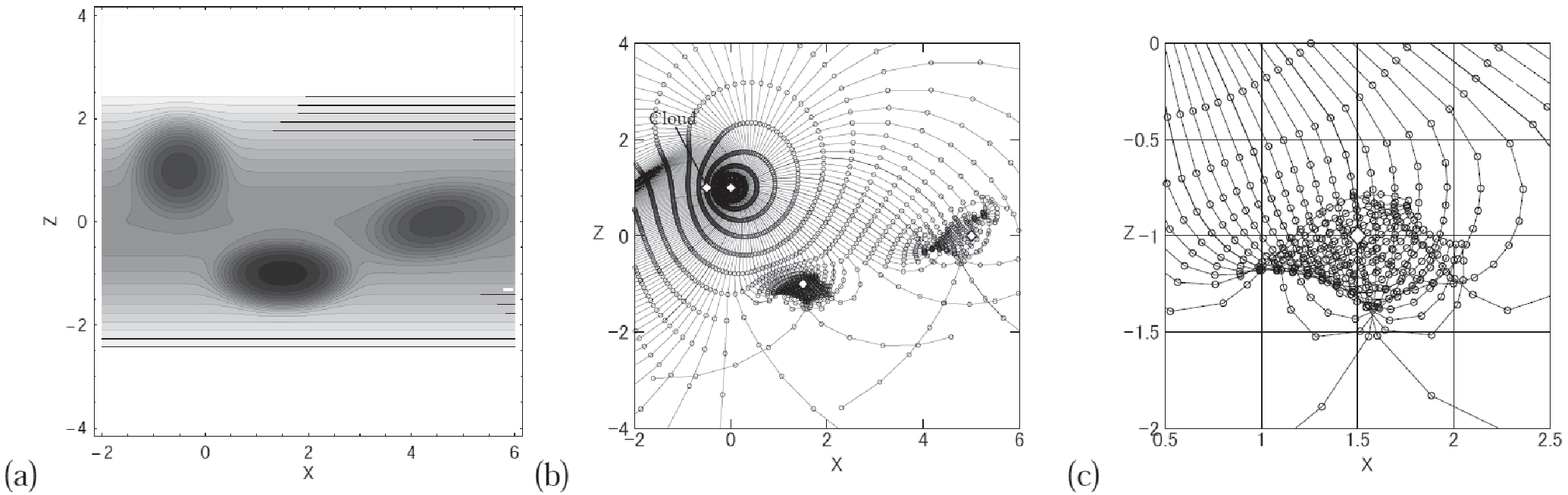} 
\end{center}
\caption{Same as figure \ref{mhdM16}, but clouds are elongated and expressed by
$\rho=0.1+\sech(z/1.0) 
         +10 \exp(-(((z-1)/0.5)^2+((x+0.2)/0.5)^2)
         +50 \exp(-(((z+1)/0.3)^2+((x-1.5)/0.6)^2)
         +10 \exp(-((z/0.4)^2+((-0.5z+x-5)/0.7)^2)).$ }
\label{mhdM16_elong} 

\begin{center}   
\includegraphics[width=14cm]{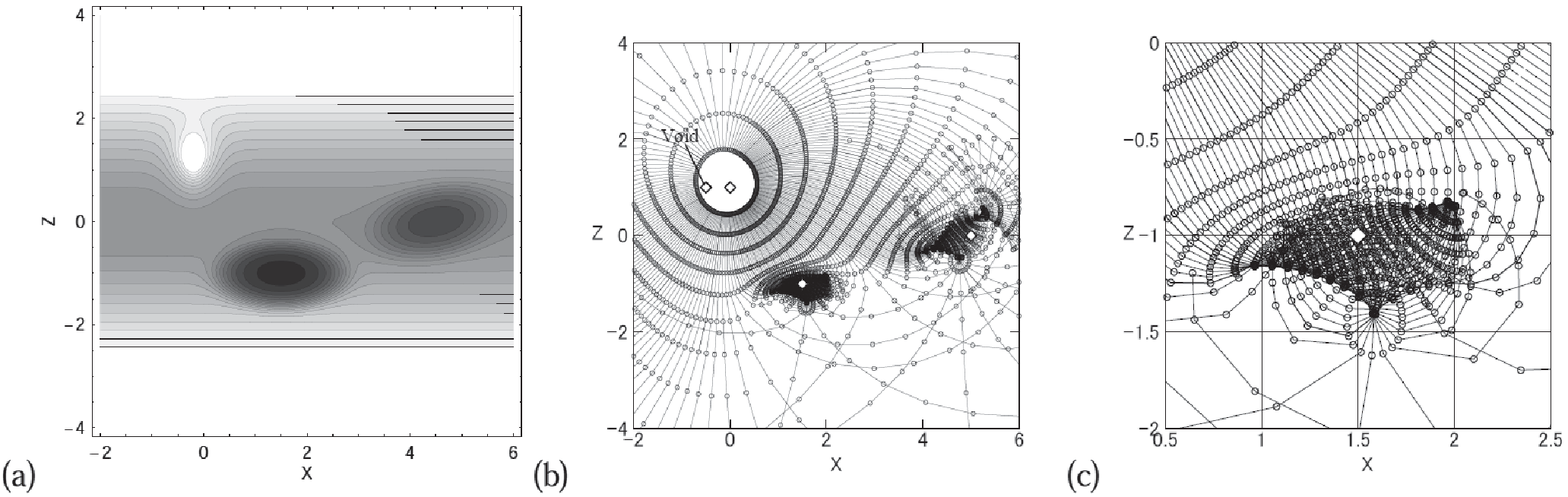} 
\end{center}
\caption{
{\revise Same as Fig. \ref{mhdM16_elong}, but the nearest cloud is replaced with a void of amplitude $-0.6$. Note the faster and higher-efficiency convergence of waves onto the other clouds than in figure \ref{mhdM16_elong}.}}
\label{void_mhdM16_elong} 
	\end{figure*} 

Figure \ref{M16M17}(a) shows \co-line map at $\vlsr\sim 20$ \kms around M16 and M17 made from the CO survey data cube observed with the Nobeyama-45 m telescope (Umemoto et al. 2017). The CO map reveals a long string of molecular clouds from M16 to the GMC at G12-14 with M17 being located in touch with the dense GMC at G14.7-0.5. Adopting a spectroscopic distance of $2.1\pm 0.2$ kpc to M17 (Heffmeister et al. 2008), the projected extent of cloud string from G12 to G18 is as long as $\sim 200$ pc.   

Figure \ref{M16M17}(b) shows a radio continuum map of a wider area around NGC 6604, M16 and M17 at 2.7 GHz from the galactic plane survey observed with the Effelsberg-100 m telescope (Reich et al. 1984), where the large-scale background emission has been subtracted. The giant radio shells identified by Sofue et al. (1986) are indicated by the three dashed ellipses in touch with M16. Also indicated by longer dashed curves are possible outer shells, which might be older and is almost reaching M17.

Figure \ref{m17_coshells} shows a close up view the \co map around M17 at 20 \kms with an overlaid high-resolution radio continuum map of the HII region at 90 by contours. The HII region grows from the center of a CO shell, blistering toward the south in an open cone. The right panel enlarges the CO line shell in the CO line at 15.4\kms, revealing a half-shell structure.

The remote triggering scenario for M16 and M17 is now simulated by putting three Gaussian clouds as shown in figure \ref{mhdM16}(a), mimicking the CO clouds in figure \ref{M16M17}. A point explosion is assumed to occur at $(x,y)=(0,1)$ at the western edge of the first cloud centered on $(x,y)=(-0.5,1)$, mimicking a blast wave from M16. As the wave front (right panel) expands, a significant fraction converges to the focal point of the 2nd cloud on the opposite side of the cloud center. 

The same calculation was also performed for elongated clouds as shown in figure \ref{mhdM16_elong}, where the M17 cloud is elongated in the $x$ direction and G013 cloud is elongated and inclined from the galactic plane. The wave propagation and convergence are essentially the same as for the spherical clouds, except that the slightly worse efficiency of focusing. 
{\revise Also a case that the nearest cloud is replaced with a void, which mimics a case that the surrounding gas has been swept away by the expanding HII shell of M16, is examined in figure \ref{void_mhdM16_elong}.}  
  
In the case for spherical clouds, the waves sharply converge to the focal point of M17 cloud, where the wave amplitude increases drastically (figure \ref{flux}). It is interesting to notice that a high-density spherical shell of molecular gas is indeed observed around M17 in the CO-line channel maps, as shown in figure \ref{m17_coshells}. Such a concentric CO shell could be formed by late-arrival of MHD waves emitted intermittently during the breathing SF activity in M16. 

 An HII region W33, whose distance is 2.4 kpc and radial velocity of $\sim 35$ \kms, marked also by the asterisque, could be a member of the long chain of HII regions continuing from NGC 6604 (W35). It is interesting to point out that the simulation shows another FMW on the 3rd cloud, suggesting that W33 has also a common origin to M17 and M16. 

{\revise However, the convergence to M17 is less effective for the case with elongated clouds. On the other hand, if the nearest cloud is replaced by a cavity, the convergence to M17 and W35 becomes much more effective, which is due to the  reflection of considerable portion of the released wave to the other side facing the other clouds.}

The result may be compared with other interpretations.  Hoffmeister et al. (2008) argues for on-going SF in M17 triggered by the central O stars along the sequential SF scenario, while the origin of the O stars must be explained by other triggering mechanism. Nishimura et al. (2018) suggest collision scenario, attributing a high-velocity CO component to collision without determining the cloud's orbit. Thus, the present FMW model would be worth being considered as an alternative or complimentary mechanism of trigger of SF in M17.

\section{Discussion} 
 
\sub{ High-efficiency feedback}  
The fast-mode MHD waves focus not only on nearby clouds, but also on distant clouds from the wave source. Significant fraction of the released waves converge onto the clouds and are gathered to the focal points, realizing high-efficiency feedback of the energy to the surrounding clouds.

Even an isolated cloud without any nearby SF activity suffers from invasion of waves from a distant SF site. The high rate of reflection coefficient by the increasing \Alf velocity toward the halo guarantees the confinement of the waves inside the galactic disc, so that most of the released waves are confined to the disc and converge onto the molecular clouds.

{\revise In a filamentary cloud, wave focusing takes place periodically along the major axis at high convergence efficiency. This suggests that SF sites appear with periodicity along the major axis at an interval comparable to the filament's width.

A void, or cavity, of ISM penetrated by magnetic fields reflects MHD waves, handing the wave energy more quickly to the other clouds. This means that the focusing SF is more efficient, if the interstellar space is filled with cavities.}
 
\sub{ Spherical implosion}
An advantageous point of the FMW is the 3D geometrical effect during the spherical implosion. This realizes a much higher compression onto the focal point compared to 2D shock compression in a sheet as postulated for the sequential SF and cloud-collision.

Suppose that about $\sim 10$\% of the released waves at the source is converging to the focal point, as in figure \ref{mhdM16}. Then, the wave amplitude increases inversely proportional to the front's surface. In the simulation of M16 and M17, the flux may be approximately calculated as 
$ F\sim \alpha \epsilon_{\rm HII}V (r_{\rm HII}/ r)^2\sim  1.4\times 10^{-2}\ {\rm erg\ cm^{-2}\ s^{-1}}, $  
where 
$\epsilon_{\rm HII} \sim n_{\rm e}kT$  and $r_{\rm HII}$
 are the energy density and radius of the HII region (M16) when it was most active, respectively, with $n_{\rm e}\sim 100$ cm$^{-3}$, $T\sim 10^4$ K, and $k$ being the Boltzman constant. The efficiency was taken to be $\alpha\sim 0.1$, and $r\sim 0.1$ pc is the radius of the wave front approaching the focal point (M17) at $V\sim 1$ \kms. 

This flux may be compared with the original flux in M16, when the wave was emitted, which is on the order of 
$
F_{\rm HII}\sim \epsilon_{\rm HII}V_{\rm HII}\sim 1.4 \times 10^{-4} \ {\rm erg\ cm^{-2}\ s^{-1}}.
$
It may be also compared with the flux conveyed by a cloud of density $\rho\sim 10^3$ \Hcc at collision velocity of $v\sim 10$ \kms, which is on the order of
$
F_{\rm col}\sim \rho v^2 v/2 \sim 8\times 10^{-4}\ {\rm erg\ cm^{-2}\ s^{-1}}.
$

Thus, the energy flux at the wave front due to the 'implosive' compression in the FMW scenario is an order of magnitude greater than those due to the direct compression by HII expansion and cloud-cloud collision that are essentially 'explosive' phenomena. 

\sub{ Long-range propagation} 
The present method solves the Eikonal equations to trace the ray path, where the wave amplitude is assumed to be small. The waves can propagate over long distances up to the dissipation length, which is sufficiently longer than the distances between interstellar clouds as shown below. This will result in propagation of SF at high efficiency repeating pin-point implosion of waves from cloud to cloud.

The dissipation rate $\gamma$ of a small-amplitude MHD wave defined through amplitude $\propto \exp(-\gamma L)$ (Landau and Lifshits 1960) is expressed as
$
\gamma={(\omega^2 / 2V^3)}\({\nu/\rho} + {c^2/4\pi\sigma_e}\),
$
where $\omega$ is the frequency,  $\nu\sim 10^{-4}$ g cm s$^{-1}$ is the viscosity of hydrogen gas, $\sigma_e$ the electric conductivity, $c$ the light velocity, and $L$ is the distance along the ray path. The first term is due to dissipation by viscous energy loss, and the second term due to Ohmic loss, which is small enough compared to the first. 
Then, the dissipation length is estimated by 
$L=1/\gamma 
\sim 0.6  {\rm kpc} \(B/\mu{\rm G} \) 
\(\rho/{\rm H\ cm^{-3}}\)^{1/2} \nonumber 
\(\nu/10^{-4} {\rm g\ cm\ s^{-1}}\)^{-1}
\(\lambda/{\rm pc} \)^2
$, 
which is on the order of $L\sim 3$ kpc in the HI gas and even longer in molecular clouds for assumed wavelength $\lambda\sim 1$ pc and magnetic strength $B\sim 5$ \muG. Therefore, the dissipation is negligible in the present circumstances. 

\sub{ Boomerang and bouncing ignition}  
As shown in figure \ref{mhd_1cloud} and \ref{3d_1cloud}, the disturbance originating at a SF site on the cloud surface propagates through the cloud itself, and focuses on the opposite side of the cloud, inferring implosion of the gas and trigger SF. The newly born SF region will, then, disturb the surrounding ISM to create a new MHD wave. This new wave expands and propagates not only toward outside, but also backward through the cloud itself, tracing exactly the same path back to the 1st wave's origin. 

This boomerang effect will continue until the cloud is exhausted, exhibiting bouncing SF ignition by exchanging the waves by one after another. It must be pointed out that such bouncing ignition occurs not only inside a cloud, but also between different clouds assisted by the focal echo mechanism. It is stressed that the remote bouncing ignition will significantly increase the efficiency of triggered SF in the Galactic disc.

\sub{  Origin and fate of MHD waves}  
The MHD waves are supposed to be excited associated with shock waves by expanding HII shells around massive-star forming sites. Large concentric thermal shells as observed in radio continuum emission around M16 were raised as a possible evidence for such sites. Similar expanding shells are observed commonly in Galactic HII regions, which may be the sources for MHD waves. 

Supernovae are also strong sources of shock waves, while their direct reach is too short and the frequency of explosion is too rare to promote SF in the Galaxy. However, all they expand and fade into sound and MHD waves in their final stage. Hence, SNRs are efficient sources of interstellar MHD waves. Any violent phenomena associated with shock waves and strong disturbances can be the sources, which includes stellar winds, galactic shock waves, and activity in the Galactic Centre. Thus, the Galaxy is full of MHD wave sources, which are confined in the galactic disc and efficiently trapped to molecular clouds, where FMW would promote star formation.

Because of the long life and almost perfect confinement to the disc, the interstellar space must be filled with MHD waves, unless they are absorbed by the molecular clouds. Suppose that kinetic energy from SNR is accumulated as MHD waves for dissipation time of $t_{\rm d}\sim L/\Va \sim 3\ {\rm kpc}/20\ {\rm km\ s^{-1}} \sim 1.5\times 10^8$ y, the interstellar energy density at present must be over $\sim 10^{50} \ R_{\rm SN} t_{\rm d}/V_{\rm G} \sim 1.6\times 10^{-10}\ {\rm erg \ cm^{-3}}$, where $V_{\rm G}\sim \pi (10\ {\rm kpc})^2 \times 100\ {\rm pc}$ is the volume of the galactic disc and $R_{\rm SN}\sim 10^{-2}$ y$^{-1}$ is the SN rate. This is obviously too large compared to the observed value of $\sim 10^{-12} {\rm erg\ cm^{-3}}$. Therefore, the input MHD energy must have been exhausted by dissipation through FMW and subsequent non-linear growth at the focal points.

\sub{  Limitation of the model} : 
The present results were obtained by solving the Eikonal equations for fast-mode MHD compression waves of small amplitude. The method cannot calculate the absolute amplitude, and hence such physical quantities as the density, pressure and temperature cannot be estimated.

We may, however, speculate that the increase in the amplitude near the focus will be followed by non-linear growth and shock compression, which causes thermalization of the waves and faster damping near the focus. 
On the other hand, if SF is successfully triggered, new MHD waves may be excited, which propagate in both directions, outward to the next clouds, and backward to original SF site. Namely, the non-linear effect acts in both directions: higher dissipation to suppress propagation, and creation of new waves targeting the next focus. The latter implies that the waves are amplified and re-emitted at the focuses.

Since the wave velocity is on the order of $\sim 10-20$ \kms, it takes $\sim 3-4$ My for the waves to reach the next clouds $\sim 100$ pc away. Inside the clouds, the wave speed is slower, and it also takes a few My for the waves to focus inside the cloud. The advantage of the MHD wave model is, therefore, not the fast propagation, but its galaxy-wide, dissipation-loss, and high-efficiency feedback. 

\section{Summary}
A new model for triggered SF -- focusing MHD-wave (FMW) mechanism -- was proposed based on the simulation of propagation of fast-mode MHD waves in the galactic disc and molecular clouds by solving the Eikonal equations. The waves were shown to be efficiently trapped by nearby as well as distant molecular clouds, and converge onto the focal points. 

It was argued that such focusing disturbance causes implosive compression of molecular gas and trigger SF. Even an isolated cloud inevitably suffers from focusing implosion of disturbance from remote SF sites. The remote ignition will further lead to bouncing ignition of SF due to focal echo mechanism among SF clouds. As an observed example, the relation of M16 and M17 was discussed using molecular and radio continuum maps. Simulation suggests that the SF in M17 was triggered by FMW from M16.

\vskip 2mm

\noindent{\bf Aknowledgements}
 
Computations were carried out at the Astronomy Data Center of the National Astronomical Observatory of Japan. The author is grateful to the anonymous referee for the constructive comments.

\end{document}